\documentclass[reprint,aip,apl,amsmath,amssymb,floatfix]{revtex4-2}

\usepackage{bm}
\usepackage{graphicx}
\usepackage{esint}
\usepackage{xspace}
\usepackage{color}
\usepackage{dcolumn}
\usepackage{bm}
\usepackage{amstext}

\begin{document}
\title{Linear magnetic susceptibility of anisotropic superconductors of cuboidal
shape}
\author{Ruslan Prozorov}
\email{prozorov@ameslab.gov}

\affiliation{Ames National Laboratory, Ames, Iowa 50011, USA}
\affiliation{Department of Physics \& Astronomy, Iowa State University, Ames, Iowa
50011, U.S.A.}

\date{submitted: 28 June 2023; revised: 15 July 2023; accepted: 23 July 2023}

\begin{abstract}
A simplified model of anisotropic magnetic susceptibility in the Meissner-London
vortex-free state of cuboidal superconducting samples is presented.
Using this model, precision measurements of the magnetic response
in three perpendicular directions of a magnetic field with respect
to primary crystal axes, can be used to extract the components of
the London penetration depth, thus enabling the evaluation of the
general superfluid density tensor, which is needed in the analysis
of the superconducting gap structure. 
\end{abstract}
\maketitle

\section{Introduction}

Magnetic measurements are some of the most common ways to analyze
magnetic and superconducting materials. While in the magnets, the
response comes from atomic and spin magnetism, in superconductors,
the magnetic response is distinctly different and non-local. It comes
from the screening currents circulating at the scale of the entire
sample. 

The problem of extracting the components of the superfluid density
tensor became crucial with the discovery of highly anisotropic high$\ensuremath{-T_{c}}$
cuprates \cite{Bednorz1986,scalapino1990high}. It was realized that
the response to an applied magnetic field involves at least two components
of the London penetration depth \cite{Basov1995,Prozorov2000a,PeregBarnea2004,Fletcher2007,Martin2010,Hossain2012}.
If the crystallographic axes are not aligned along the principal sample
faces, the problem is hard to tackle. Therefore, similarly to prior
works, we will assume that crystal lattice unit cell axes are aligned
along the normal directions to the sample facets, which is assumed
to be a rectangular parallelepiped, as shown in Fig.\ref{fig:cuboid}.
One way to extract the individual components of the London penetration
depth is to use optical methods measuring by measurement of the infrared
reflectances with polarization along the $a$, $b$, and $c-$axes
\cite{Basov1995}. However, lower frequency and even DC measurements
are needed and there one has to rely on the global response of the
entire sample and in this case measured quantity depends on at least
two components of screening current. 

In order to extract separate components of the London penetration
depth, $\lambda_{i}$, where $i=a,b,c$ it was suggested to use thin
crystals and apply magnetic field along the flat faces to avoid demagnetizing
effects and, in case of layered materials, minimize the contribution
from the inter-plane currents described by the $c-$axis penetration
depth, $\lambda_{c}$. One can double the contribution of this component
by cutting the sample in half creating two new surfaces with $c-$axis
currents. When such cut sample (both halves) is re-measured it is
straightforward to use the results of both measurements to extract
both components \cite{Fletcher2007,Martin2010}. This technique is
especially useful in highly anisotropic materials. The same cutting
(cleaving) method can be used for any other component \cite{PeregBarnea2004}.
In the most sensitive techniques, only the variation of the London
penetration depth is measured precisely and determination of the absolute
values, $\lambda_{i}\left(0\right)$, requires separate efforts \cite{Prozorov2000,PeregBarnea2004,Hossain2012}.
Often, different methods are combined to obtain the final result \cite{Hossain2012,PeregBarnea2004}.

In this contribution I describe a self-contained method to estimate
the components of the penetration depth measuring bulk thick samples.
In fact a cube-shape sample will be ideal. It has the same demagnetizing
factor of $N=0.39$ in three orthogonal directions (yes, the sum must
be equal to 1 only in ellipsoids) \cite{Demag2018} and the difference
in three measured components along each axis is solely due to the
anisotropy of the superfluid response.

We are interested in weak magnetic fields, much lower than the first
critical field, $H_{c1}$, long before vortices begin entering the
sample. The impossible-to-achieve Meissner state is established when
the magnetic induction is zero everywhere in the sample. The realistic
Meissner-London state allows some magnetic field in a thin layer called
London penetration depth. In this situation, the magnetic response
is linear and, in the semi-classical approach, is governed by the
London equation \cite{Chandrasekhar1993,Einzel2003}, 
\begin{equation}
\mathbf{j}=-\mathbb{R}\mathbf{A}\label{eq:London}
\end{equation}
that connects the supercurrent density vector, $\mathbf{j}$, and
a vector potential, $\mathbf{A}$. The response tensor, $\mathbb{R}$,
consists of two parts, diamagnetic and paramagnetic, and is determined
by the superconducting gap and the parameters of the electronic band
structure \cite{Chandrasekhar1993}. Some examples of using this theory
for different superconducting order parameters can be found in Ref.~[\onlinecite{Prozorov2006}].
Equation \ref{eq:London} takes a more familiar form when the London
penetration depth is introduced, 
\begin{equation}
\lambda_{ii}^{2}=\frac{m^{*}}{\mu_{0}n_{s}e^{2}}=\frac{1}{\mu_{0}\mathbb{R}_{ii}}\label{eq:lambda}
\end{equation}
where $m^{*}$ is the effective electron mass and $n_{s}$ is the
density of electrons in a superfluid fraction in a two-fluid model
\cite{Gorter1934,Gorter1934a,Bardeen1958,Einzel2003}. Theoretically,
it is possible to consider individual components of the supercurrent
or, equivalently, the components of the London penetration depth separately,
assuming infinite or semi-infinite samples. In practice, however,
the samples are finite, and the response always contains different
components.

\section{Anisotropic magnetic susceptibility}

The total magnetic moment of an arbitrarily shaped magnetic sample
is given by \cite{Demag2018}: 
\begin{equation}
\bm{m}=\int\left(\mu_{0}^{-1}\bm{B}\left(\bm{r}\right)-\bm{H}_{0}\right)dV\label{eq:mb}
\end{equation}
where $H_{0}$ is the applied magnetic field, and the integration
is carried over the volume that completely encloses the sample. This
equation is trivial in the infinite geometries (without demagnetization
effects), but for arbitrary samples, requires a rigorous derivation
from the general equation for the magnetic moment via the integration
of shielding currents, $\bm{m}=\int\bm{r}\times\bm{j}(\bm{r})dV/2$
\cite{Demag2018}. In fact, Eq.(\ref{eq:mb}) includes the demagnetization
effects and can be used to define the effective demagnetization factors
\cite{Demag2018}. Initial magnetic flux penetration into a $\lambda-$layer
in a superconductor is linear in $\lambda/R\ll1$. When penetrating
flux sweeps about 1/3 of the sample dimension in the penetration direction,
highly non-linear (hyperbolic) functions take over so that diverging
with temperature $\lambda\left(T\right)/R$ results in zero magnetic
susceptibility. For example, for an infinite slab of width $2w$,
$\chi=(\lambda/w)\tanh(w/\lambda)-1$. However, when $\lambda/w$
ratio is small, $\tanh(w/\lambda)\approx1$, and the magnetic susceptibility
can be quite generally written as \cite{Prozorov2021}: 
\begin{equation}
\chi\left(1-N\right)=D\frac{\lambda}{R}-1\label{eq:chi}
\end{equation}
where $D$ is the dimensionality of the flux penetration, and $N$,
is the demagnetizing factor in the direction of the applied magnetic
field. In particular, $D=1$ for an infinite slab in a parallel magnetic
field, $D=2$ for an infinite cylinder in a parallel field, and $D=3$
for a sphere). In typical sub-mm crystals, $R\sim100$$\:\mu\mathrm{m}$
~[\onlinecite{Prozorov2021}] and $\lambda\left(0\right)\sim200$ nm, therefore
$\lambda\left(0\right)/R\approx2\times10^{-3}$, a small number considering
the total variation of $\chi$ from zero to $-1$. Moreover, this
is applicable practically at all temperatures. To see that, we can
estimate the temperature at which $\lambda$ doubles. Using well-known
approximation, $\lambda\left(t\right)=\lambda_{0}/\sqrt{1-t^{2}}$
 [\onlinecite{Prozorov2006}] ($t=T/T_{c}$ - reduced temperature), we estimate
that the penetration depth doubles at $t=\sqrt{3}/2\approx0.87$,
which is practically the entire temperature range.

\begin{figure}[tb]
\includegraphics[width=8cm]{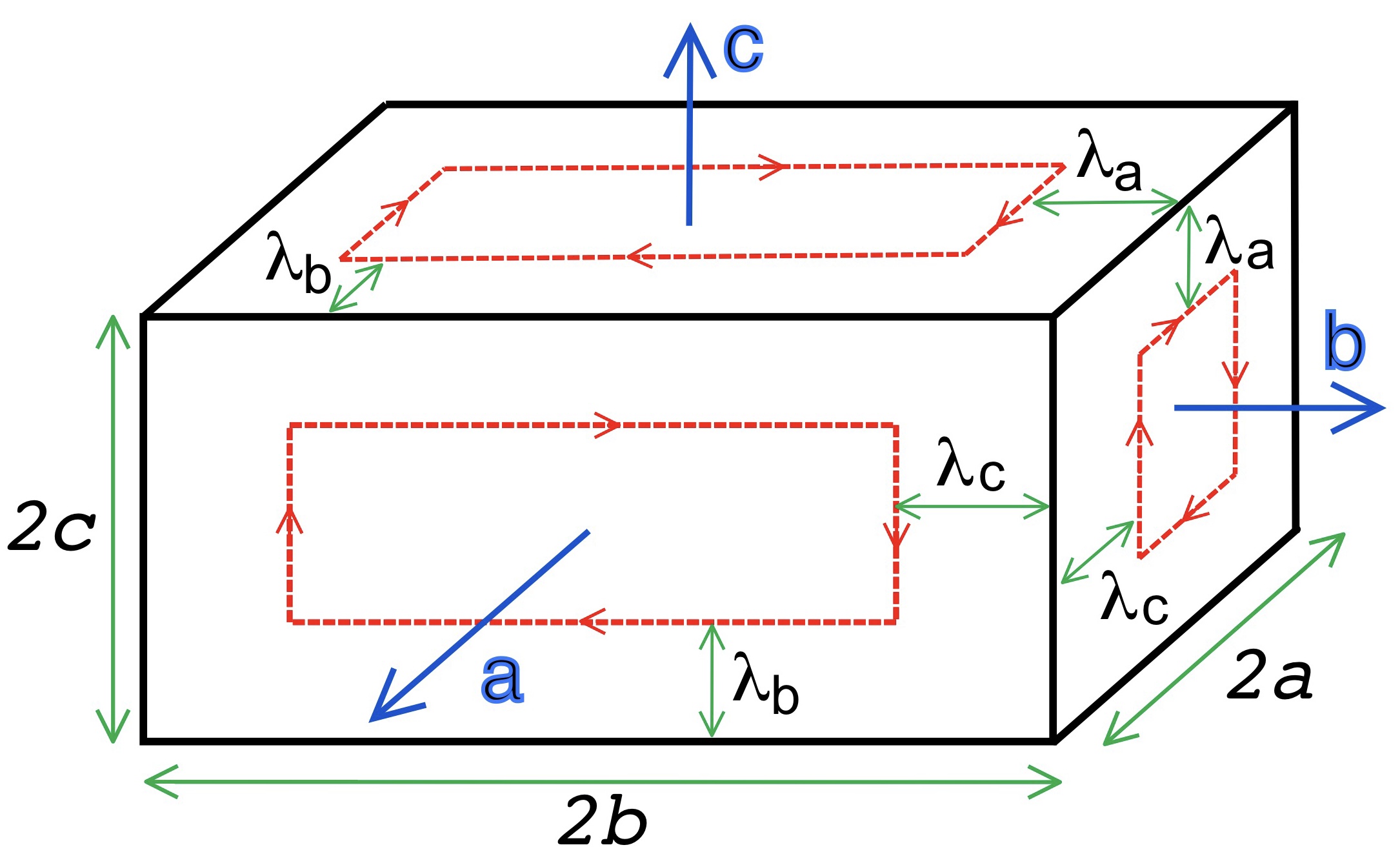} \caption{The definitions of symbols and directions in a cuboidal sample of
size $2a\times2b\times2c$ showing different components of the London
penetration depths. Directions of screening currents flow are shown
by loops with arrows.}
\label{fig:cuboid} 
\end{figure}

Linear approximation, Eq.\ref{eq:chi}, means that we can
calculate magnetic susceptibility as the ratio of the shielded volume
to the total volume. Owing to the exponential attenuation of the magnetic
field from the surface, one can roughly assume a complete field penetration
into the layer of depth $\lambda$ and no magnetic field at all beyond
that. Then, the volume penetrated by the magnetic flux is determined
by the corresponding components of the London penetration depth, as
shown in Fig.\ref{fig:cuboid}.

We need to define the components of the current density and the London
penetration depth with respect to crystal axes. Consider a cuboid-shaped
crystal of size $2a\times2b\times2c$. The vectors $\mathbf{a},\mathbf{b}$
and $\mathbf{c}$ point along the corresponding lengths as shown in 
Fig.\ref{fig:cuboid}. When a magnetic field is directed along the
$\mathbf{c}-$axis, screening currents flow in the $\mathbf{ab}-$plane.
Current flowing along the $a-$side is attenuated on the length $\lambda_{a}$,
whereas current flowing along the $b-$side by length $\lambda_{b}$.
The gradient of the magnetic induction is perpendicular to the local
direction of the current flow.

To obtain three unknown components of the London penetration depth
one needs three components of magnetic susceptibility. Assuming that
sample crystallographic directions are parallel to the sample sides,
magnetic susceptibility needs to be measured in three principal directions,
so that, for example, $\chi_{a}$ is magnetic susceptibility measured
with a magnetic field applied along the $\mathbf{a}-$axis. For the
analysis, it is crucially important to obtain properly normalized
magnetic susceptibilities so that they include (direction-dependent)
demagnetization correction, $\left(1-N\right),$ Using well-knowsee Eq.\ref{eq:chi}.
For example, we should have $\chi_{a}=-1$ when $\lambda_{c}=\lambda_{b}=0$
(ideal shielding). The practical experimental procedure depends on
the measurement technique and measured quantity.

In the case of a DC magnetometry (Quantum Design MPMS, VSM, extraction
magnetometer, Faraday balance, etc.), one can take the value of the
magnetic moment at the lowest temperature and use it for normalization.
For example, if a total magnetic moment, $m(T)$, is measured, then
the normalized magnetic susceptibility is given my $\chi(T)=m(T)/|m_{0}|$,
where the denominator is the theoretical magnetic moment of a perfect
diamagnetic sample of the same shape, $m_{0}=-VH_{0}/(1-N)$, $V$
is sample volume and $H_{0}$ is the applied magnetic field. The normalization
is performed assuming $m_{0}\approx m\left(T_{min}\right)$. Unfortunately,
insufficient sensitivity, limited dynamic range, and omnipresent noise
make it practically impossible to use DC magnetometry to determine
the London penetration depth. 

The biggest problem is that there is no true reference point, ideal
diamagnetic response (i.e., a pure Meissner state), so the measurements
are always performed with finite $\lambda(T)$, and one does not know
what the signal would be with $\lambda=0$. Usually, only the variation
of the London penetration depth with temperature is measured since
it contains information about low-energy quasiparticles, hence the
order parameter structure. For example, the rate of change is about
5 \AA/K in YBCO, and 20 \AA/K in BSCCO \cite{Basov1995,Prozorov2000,Hossain2012}. 

Let us take a superconducting ball of radius $R$. Its magnetic susceptibility
in the pure Meissner state is $\chi=-\left(3/2\right)V=-2\pi R^{3}$,
where $3/2=1/(1-1/3)$ prefactor is the demagnetization correction.
Suppose we would like to detect the 5 \AA\, change of $\Delta\lambda\left(T\right)$
warming the sample by 1 K from the base temperature. The corresponding volume change
in the ball is $4\pi R^{2}\Delta\lambda$. Therefore, we need to resolve
the change of magnetic susceptibility, $4\pi R^{2}\Delta\lambda/2\pi R^{3}=2\Delta\lambda/R$.
For a mm-sized crystal, we have for YBCO, $2\Delta\lambda/R\approx10^{-6}$.
The total magnetic moment of 1 mm ball is $m=-2\pi R^{3}H\approx-6\times10^{-3}H$
emu, where magnetic field is in Oe. Therefore, one needs to detect
a change in the magnetic moment of $\Delta m\approx6\times10^{-9}H$
emu. Here quoted sensitivity of the magnetometers mentioned above
is $10^{-6}\;\text{emu}$. Therefore, with the excitation fields less
than 10 Oe (usually much less), the required sensitivity of better
than $6\times10^{-8}$ emu is far beyond these instruments capabilities. Note that, ideally, we would want to resolve at least ten points in that 1 K interval. Then the resolution needs to be 10 time better. Of course, the magnetic moment is larger in higher fields, but one needs to worry about vortices at some point. Of course, this reasoning only applies
to commercial systems measuring total magnetic moment of a sample.

\begin{figure}[tb]
\includegraphics[width=8cm]{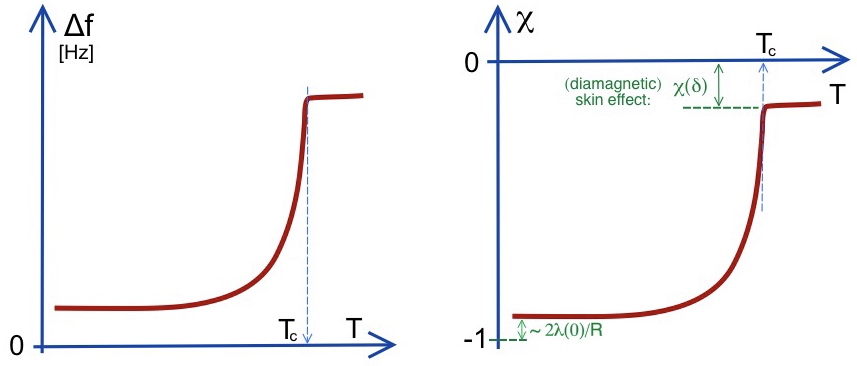} \caption{(left panel) Schematics of the measured signal, shown as the resonant
frequency shift in the case of the frequency-domain susceptometry;
(right panel) calibrated effective susceptibility. The deviations
from the ideal behavior are associated with the finite London penetration
depth at low temperatures and skin depth near $T_{c}$.}
\label{fig:measurement} 
\end{figure}

In the case of frequency-domain AC susceptibility measurements, such
as tunnel-diode resonator (TDR) \cite{Carrington2011,Giannetta2022,Prozorov2006,Prozorov2011},
microwave cavity perturbation \cite{Donovan1993,Dressel1993,Klein1993}
or even amplitude-domain conventional AC susceptibility \cite{Martin1995,Topping2018},
sufficient sensitivity and dynamic range can be achieved (although
this requires a significant effort) for precision measurements of
$\lambda(T)$. However, one can no longer take the signal difference
from above $T_{c}$ to the lowest temperature because of the screening
of the applied AC fields due to the normal skin effect. This is shown
schematically in Fig.\ref{fig:measurement} where the left panel shows
typical resonant frequency variation upon sweeping temperature through
$T_{c}$. The calibrated magnetic susceptibility needed for the analysis
is depicted in the right panel. To determine the total screening,
one can estimate the normal-state screening from the independent resistivity
measurements through $T_{c}$. This is not the most precise approach
because it involves sample-dependent calibration factors. Alternatively,
the measurement device could allow for the physical extraction of
the sample from the sensing coil \textit{in situ}. This requires a
substantial modification of the experimental setup, but once built,
it gives the ability to calibrate every measured sample and automatically
includes the demagnetization correction. This is implemented, for
example, in our tunnel-diode resonators \cite{Prozorov2011,Prozorov2006,Giannetta2022}.
At low temperatures, the deviation from perfect screening, $2\lambda\left(0\right)/R\approx4\times10^{-3}$,
is a small number.

\section{London penetration depth}

Assuming that we have all three components of the normalized magnetic
susceptibility, it is convenient to construct a vector of ``normalized
penetrated volumes'',

\begin{equation}
\bm{X}=\left(1+\chi_{a}\left(T\right),1+\chi_{b}\left(T\right),1+\chi_{c}\left(T\right)\right)
\end{equation}
whose components vary from $1$ in case of a complete penetration
($\chi_{i}=0$) to $0$ in case of a complete screening ($\chi_{i}=-1$).
For example, for $\mathbf{H}||\mathbf{c}$, the normalized penetrated volume is: 
\begin{equation}
\chi_{c}=\frac{(2c)(2a)(2\lambda_{a})+(2c)(2b)(2\lambda_{b})}{V}=\frac{\lambda_{a}}{b}+\frac{\lambda_{b}}{a}
\end{equation}
where sample volume $V=(2a)(2b)(2c)=8abc$. Doing the same for other two orientations, we obtain: 
\begin{equation}
\bm{X}=\bm{L}\cdot\bm{\Lambda}
\end{equation}
where the London penetration depth vector $\bm{\Lambda}=\left(\lambda_{a},\lambda_{b},\lambda_{c}\right)$
and the coupling London matrix is: 
\begin{equation}
\bm{L}=\left(\begin{array}{ccc}
0 & \frac{1}{c} & \frac{1}{b}\\
\frac{1}{c} & 0 & \frac{1}{a}\\
\frac{1}{b} & \frac{1}{a} & 0
\end{array}\right)
\end{equation}
The solution of this vector equation is, 
\begin{equation}
\bm{\Lambda}=\bm{L}^{-1}\cdot\bm{X}
\end{equation}
where the inverse of the London matrix is given by: 
\begin{equation}
\bm{L}^{-1}\bm{=}\left(\begin{array}{ccc}
-\frac{bc}{a} & c & b\\
c & -\frac{ac}{b} & a\\
b & a & -\frac{ab}{c}
\end{array}\right)
\end{equation}
The resulting solution written in components is: 
\begin{equation}
\left\{ \begin{array}{l}
2\lambda_{a}=-\frac{bc}{a}\left(1+\chi_{a}\right)+c\left(1+\chi_{b}\right)+b\left(1+\chi_{c}\right)\\
2\lambda_{b}=c\left(1+\chi_{a}\right)-\frac{ac}{b}\left(1+\chi_{b}\right)+a\left(1+\chi_{c}\right)\\
2\lambda_{c}=b\left(1+\chi_{a}\right)+a\left(1+\chi_{b}\right)-\frac{ab}{c}\left(1+\chi_{c}\right)
\end{array}\right.\label{eq:lambdas-full}
\end{equation}
This allows evaluating three principal components of the London penetration
depths from three independent magnetic susceptibility measurements.
Indeed, it is hard to find perfect samples with ideal geometry, so
errors in the amplitudes are inevitable. However, this procedure may
help identify some non-trivial temperature dependencies of the penetration
depth, distinguish nodal from nodeless states, or correlate with resistivity
anisotropies in search of nematic states. These estimates may be useful
when other parameters are also available, for example, specific heat,
resistivity, and upper critical field, which are tied together by
thermodynamic Rutgers relations \cite{Rutgers1934,Kim2013}.

\section{tetragonal crystals}

\noindent The general relation, Eq.\ref{eq:lambdas-full}, is simplified
if one considers a quite typical for superconductors case of tetragonal
(or close to tetragonal) symmetry of the crystal lattice. High$-T_{c}$
cuprates and iron-based superconductors are notable examples. In this
case, there are two principal values of the London penetration depth,
in plane, $\lambda_{a}=\lambda_{b}=\lambda_{ab}$, and out of plane,
$\lambda_{c}.$ Note, however, that the sample still has three different
sides, $a,b,c$. This means that all three components of the magnetic
susceptibility will be different. In this case Eq.\ref{eq:lambdas-full}
is simplified as:

\begin{equation}
\left\{ \begin{array}{l}
\lambda_{ab}=\frac{ab}{a+b}\left(1+\chi_{c}\right)\\
\lambda_{c}=b\left(1+\chi_{a}\right)-\frac{ab^{2}}{c\left(a+b\right)}\left(1+\chi_{c}\right)\\
....=a\left(1+\chi_{b}\right)-\frac{ba^{2}}{c\left(a+b\right)}\left(1+\chi_{c}\right)\\
....=\frac{ab}{b-a}\left(\chi_{b}-\chi_{a}\right)
\end{array}\right.
\end{equation}

These are useful formulas as they show that in order to evaluate the
in-plane London penetration depth one needs to measure only $\chi_{c}$
which is what most experimentalists do. To obtain the $c-$axis penetration
depth one needs to measure perpendicular components $\chi_{a}$ and/or
$\chi_{b}$. Having both will improve the accuracy of the estimate.
\begin{acknowledgments}
I thank Takasada Shibauchi and Kota Ishihara for constructive discussions.
This work was supported by the U.S. Department of Energy (DOE), Office
of Science, Basic Energy Sciences, Materials Science and Engineering
Division. Ames Laboratory is operated for the U.S. DOE by Iowa State
University under contract DE-AC02-07CH11358. 
\end{acknowledgments}


\begin{thebibliography}{27}%
\makeatletter
\providecommand \@ifxundefined [1]{%
 \@ifx{#1\undefined}
}%
\providecommand \@ifnum [1]{%
 \ifnum #1\expandafter \@firstoftwo
 \else \expandafter \@secondoftwo
 \fi
}%
\providecommand \@ifx [1]{%
 \ifx #1\expandafter \@firstoftwo
 \else \expandafter \@secondoftwo
 \fi
}%
\providecommand \natexlab [1]{#1}%
\providecommand \enquote  [1]{``#1''}%
\providecommand \bibnamefont  [1]{#1}%
\providecommand \bibfnamefont [1]{#1}%
\providecommand \citenamefont [1]{#1}%
\providecommand \href@noop [0]{\@secondoftwo}%
\providecommand \href [0]{\begingroup \@sanitize@url \@href}%
\providecommand \@href[1]{\@@startlink{#1}\@@href}%
\providecommand \@@href[1]{\endgroup#1\@@endlink}%
\providecommand \@sanitize@url [0]{\catcode `\\12\catcode `\$12\catcode
  `\&12\catcode `\#12\catcode `\^12\catcode `\_12\catcode `\%12\relax}%
\providecommand \@@startlink[1]{}%
\providecommand \@@endlink[0]{}%
\providecommand \url  [0]{\begingroup\@sanitize@url \@url }%
\providecommand \@url [1]{\endgroup\@href {#1}{\urlprefix }}%
\providecommand \urlprefix  [0]{URL }%
\providecommand \Eprint [0]{\href }%
\providecommand \doibase [0]{https://doi.org/}%
\providecommand \selectlanguage [0]{\@gobble}%
\providecommand \bibinfo  [0]{\@secondoftwo}%
\providecommand \bibfield  [0]{\@secondoftwo}%
\providecommand \translation [1]{[#1]}%
\providecommand \BibitemOpen [0]{}%
\providecommand \bibitemStop [0]{}%
\providecommand \bibitemNoStop [0]{.\EOS\space}%
\providecommand \EOS [0]{\spacefactor3000\relax}%
\providecommand \BibitemShut  [1]{\csname bibitem#1\endcsname}%
\let\auto@bib@innerbib\@empty
\bibitem [{\citenamefont {Bednorz}\ and\ \citenamefont
  {M\"uller}(1986)}]{Bednorz1986}%
  \BibitemOpen
  \bibfield  {author} {\bibinfo {author} {\bibfnamefont {J.~G.}\ \bibnamefont
  {Bednorz}}\ and\ \bibinfo {author} {\bibfnamefont {K.~A.}\ \bibnamefont
  {M\"uller}},\ }\href {https://doi.org/10.1007/BF0130370} {\bibfield
  {journal} {\bibinfo  {journal} {Zeitschrift f\"ur Physik B Condensed Matter}\
  }\textbf {\bibinfo {volume} {64}},\ \bibinfo {pages} {189} (\bibinfo {year}
  {1986})}\BibitemShut {NoStop}%
\bibitem [{\citenamefont {Scalapino}(1990)}]{scalapino1990high}%
  \BibitemOpen
  \bibfield  {author} {\bibinfo {author} {\bibfnamefont {D.}~\bibnamefont
  {Scalapino}},\ }in\ \href@noop {} {\emph {\bibinfo {booktitle} {The Los
  Alamos Symposium}}}\ (\bibinfo {organization} {Addison-Wesley},\ \bibinfo
  {year} {1990})\BibitemShut {NoStop}%
\bibitem [{\citenamefont {Basov}\ \emph {et~al.}(1995)\citenamefont {Basov},
  \citenamefont {Liang}, \citenamefont {Bonn}, \citenamefont {Hardy},
  \citenamefont {Dabrowski}, \citenamefont {Quijada}, \citenamefont {Tanner},
  \citenamefont {Rice}, \citenamefont {Ginsberg},\ and\ \citenamefont
  {Timusk}}]{Basov1995}%
  \BibitemOpen
  \bibfield  {author} {\bibinfo {author} {\bibfnamefont {D.~N.}\ \bibnamefont
  {Basov}}, \bibinfo {author} {\bibfnamefont {R.}~\bibnamefont {Liang}},
  \bibinfo {author} {\bibfnamefont {D.~A.}\ \bibnamefont {Bonn}}, \bibinfo
  {author} {\bibfnamefont {W.~N.}\ \bibnamefont {Hardy}}, \bibinfo {author}
  {\bibfnamefont {B.}~\bibnamefont {Dabrowski}}, \bibinfo {author}
  {\bibfnamefont {M.}~\bibnamefont {Quijada}}, \bibinfo {author} {\bibfnamefont
  {D.~B.}\ \bibnamefont {Tanner}}, \bibinfo {author} {\bibfnamefont {J.~P.}\
  \bibnamefont {Rice}}, \bibinfo {author} {\bibfnamefont {D.~M.}\ \bibnamefont
  {Ginsberg}},\ and\ \bibinfo {author} {\bibfnamefont {T.}~\bibnamefont
  {Timusk}},\ }\href {https://doi.org/10.1103/PhysRevLett.74.598} {\bibfield
  {journal} {\bibinfo  {journal} {Phys. Rev. Lett.}\ }\textbf {\bibinfo
  {volume} {74}},\ \bibinfo {pages} {598} (\bibinfo {year} {1995})}\BibitemShut
  {NoStop}%
\bibitem [{\citenamefont {Prozorov}\ \emph
  {et~al.}(2000{\natexlab{a}})\citenamefont {Prozorov}, \citenamefont
  {Giannetta}, \citenamefont {Carrington},\ and\ \citenamefont
  {Araujo-Moreira}}]{Prozorov2000a}%
  \BibitemOpen
  \bibfield  {author} {\bibinfo {author} {\bibfnamefont {R.}~\bibnamefont
  {Prozorov}}, \bibinfo {author} {\bibfnamefont {R.~W.}\ \bibnamefont
  {Giannetta}}, \bibinfo {author} {\bibfnamefont {A.}~\bibnamefont
  {Carrington}},\ and\ \bibinfo {author} {\bibfnamefont {F.~M.}\ \bibnamefont
  {Araujo-Moreira}},\ }\href {<Go to ISI>://WOS:000088037000030
  http://prb.aps.org/pdf/PRB/v62/i1/p115_1} {\bibfield  {journal} {\bibinfo
  {journal} {Phys. Rev. B}\ }\textbf {\bibinfo {volume} {62}},\ \bibinfo
  {pages} {115} (\bibinfo {year} {2000}{\natexlab{a}})}\BibitemShut {NoStop}%
\bibitem [{\citenamefont {Pereg-Barnea}\ \emph {et~al.}(2004)\citenamefont
  {Pereg-Barnea}, \citenamefont {Turner}, \citenamefont {Harris}, \citenamefont
  {Mullins}, \citenamefont {Bobowski}, \citenamefont {Raudsepp}, \citenamefont
  {Liang}, \citenamefont {Bonn},\ and\ \citenamefont
  {Hardy}}]{PeregBarnea2004}%
  \BibitemOpen
  \bibfield  {author} {\bibinfo {author} {\bibfnamefont {T.}~\bibnamefont
  {Pereg-Barnea}}, \bibinfo {author} {\bibfnamefont {P.~J.}\ \bibnamefont
  {Turner}}, \bibinfo {author} {\bibfnamefont {R.}~\bibnamefont {Harris}},
  \bibinfo {author} {\bibfnamefont {G.~K.}\ \bibnamefont {Mullins}}, \bibinfo
  {author} {\bibfnamefont {J.~S.}\ \bibnamefont {Bobowski}}, \bibinfo {author}
  {\bibfnamefont {M.}~\bibnamefont {Raudsepp}}, \bibinfo {author}
  {\bibfnamefont {R.}~\bibnamefont {Liang}}, \bibinfo {author} {\bibfnamefont
  {D.~A.}\ \bibnamefont {Bonn}},\ and\ \bibinfo {author} {\bibfnamefont
  {W.~N.}\ \bibnamefont {Hardy}},\ }\href
  {https://doi.org/10.1103/physrevb.69.184513} {\bibfield  {journal} {\bibinfo
  {journal} {Physical Review B}\ }\textbf {\bibinfo {volume} {69}},\ \bibinfo
  {pages} {184513} (\bibinfo {year} {2004})}\BibitemShut {NoStop}%
\bibitem [{\citenamefont {Fletcher}\ \emph {et~al.}(2007)\citenamefont
  {Fletcher}, \citenamefont {Carrington}, \citenamefont {Diener}, \citenamefont
  {Rodi{\`{e}}re}, \citenamefont {Brison}, \citenamefont {Prozorov},
  \citenamefont {Olheiser},\ and\ \citenamefont {Giannetta}}]{Fletcher2007}%
  \BibitemOpen
  \bibfield  {author} {\bibinfo {author} {\bibfnamefont {J.~D.}\ \bibnamefont
  {Fletcher}}, \bibinfo {author} {\bibfnamefont {A.}~\bibnamefont
  {Carrington}}, \bibinfo {author} {\bibfnamefont {P.}~\bibnamefont {Diener}},
  \bibinfo {author} {\bibfnamefont {P.}~\bibnamefont {Rodi{\`{e}}re}}, \bibinfo
  {author} {\bibfnamefont {J.~P.}\ \bibnamefont {Brison}}, \bibinfo {author}
  {\bibfnamefont {R.}~\bibnamefont {Prozorov}}, \bibinfo {author}
  {\bibfnamefont {T.}~\bibnamefont {Olheiser}},\ and\ \bibinfo {author}
  {\bibfnamefont {R.~W.}\ \bibnamefont {Giannetta}},\ }\href
  {https://doi.org/10.1103/physrevlett.98.057003} {\bibfield  {journal}
  {\bibinfo  {journal} {Physical Review Letters}\ }\textbf {\bibinfo {volume}
  {98}},\ \bibinfo {pages} {057003} (\bibinfo {year} {2007})}\BibitemShut
  {NoStop}%
\bibitem [{\citenamefont {Martin}\ \emph {et~al.}(2010)\citenamefont {Martin},
  \citenamefont {Kim}, \citenamefont {Gordon}, \citenamefont {Ni},
  \citenamefont {Kogan}, \citenamefont {Bud'ko}, \citenamefont {Canfield},
  \citenamefont {Tanatar},\ and\ \citenamefont {Prozorov}}]{Martin2010}%
  \BibitemOpen
  \bibfield  {author} {\bibinfo {author} {\bibfnamefont {C.}~\bibnamefont
  {Martin}}, \bibinfo {author} {\bibfnamefont {H.}~\bibnamefont {Kim}},
  \bibinfo {author} {\bibfnamefont {R.~T.}\ \bibnamefont {Gordon}}, \bibinfo
  {author} {\bibfnamefont {N.}~\bibnamefont {Ni}}, \bibinfo {author}
  {\bibfnamefont {V.~G.}\ \bibnamefont {Kogan}}, \bibinfo {author}
  {\bibfnamefont {S.~L.}\ \bibnamefont {Bud'ko}}, \bibinfo {author}
  {\bibfnamefont {P.~C.}\ \bibnamefont {Canfield}}, \bibinfo {author}
  {\bibfnamefont {M.~A.}\ \bibnamefont {Tanatar}},\ and\ \bibinfo {author}
  {\bibfnamefont {R.}~\bibnamefont {Prozorov}},\ }\href
  {https://doi.org/10.1103/physrevb.81.060505} {\bibfield  {journal} {\bibinfo
  {journal} {Physical Review B}\ }\textbf {\bibinfo {volume} {81}},\ \bibinfo
  {pages} {060505} (\bibinfo {year} {2010})}\BibitemShut {NoStop}%
\bibitem [{\citenamefont {Hossain}\ \emph {et~al.}(2012)\citenamefont
  {Hossain}, \citenamefont {Baglo}, \citenamefont {Wojek}, \citenamefont
  {Ofer}, \citenamefont {Dunsiger}, \citenamefont {Morris}, \citenamefont
  {Prokscha}, \citenamefont {Saadaoui}, \citenamefont {Salman}, \citenamefont
  {Bonn}, \citenamefont {Liang}, \citenamefont {Hardy}, \citenamefont {Suter},
  \citenamefont {Morenzoni},\ and\ \citenamefont {Kiefl}}]{Hossain2012}%
  \BibitemOpen
  \bibfield  {author} {\bibinfo {author} {\bibfnamefont {M.}~\bibnamefont
  {Hossain}}, \bibinfo {author} {\bibfnamefont {J.}~\bibnamefont {Baglo}},
  \bibinfo {author} {\bibfnamefont {B.}~\bibnamefont {Wojek}}, \bibinfo
  {author} {\bibfnamefont {O.}~\bibnamefont {Ofer}}, \bibinfo {author}
  {\bibfnamefont {S.}~\bibnamefont {Dunsiger}}, \bibinfo {author}
  {\bibfnamefont {G.}~\bibnamefont {Morris}}, \bibinfo {author} {\bibfnamefont
  {T.}~\bibnamefont {Prokscha}}, \bibinfo {author} {\bibfnamefont
  {H.}~\bibnamefont {Saadaoui}}, \bibinfo {author} {\bibfnamefont
  {Z.}~\bibnamefont {Salman}}, \bibinfo {author} {\bibfnamefont
  {D.}~\bibnamefont {Bonn}}, \bibinfo {author} {\bibfnamefont {R.}~\bibnamefont
  {Liang}}, \bibinfo {author} {\bibfnamefont {W.}~\bibnamefont {Hardy}},
  \bibinfo {author} {\bibfnamefont {A.}~\bibnamefont {Suter}}, \bibinfo
  {author} {\bibfnamefont {E.}~\bibnamefont {Morenzoni}},\ and\ \bibinfo
  {author} {\bibfnamefont {R.}~\bibnamefont {Kiefl}},\ }\href
  {https://doi.org/10.1016/j.phpro.2012.04.081} {\bibfield  {journal} {\bibinfo
   {journal} {Physics Procedia}\ }\textbf {\bibinfo {volume} {30}},\ \bibinfo
  {pages} {235} (\bibinfo {year} {2012})}\BibitemShut {NoStop}%
\bibitem [{\citenamefont {Prozorov}\ \emph
  {et~al.}(2000{\natexlab{b}})\citenamefont {Prozorov}, \citenamefont
  {Giannetta}, \citenamefont {Carrington}, \citenamefont {Fournier},
  \citenamefont {Greene}, \citenamefont {Guptasarma}, \citenamefont {Hinks},\
  and\ \citenamefont {Banks}}]{Prozorov2000}%
  \BibitemOpen
  \bibfield  {author} {\bibinfo {author} {\bibfnamefont {R.}~\bibnamefont
  {Prozorov}}, \bibinfo {author} {\bibfnamefont {R.~W.}\ \bibnamefont
  {Giannetta}}, \bibinfo {author} {\bibfnamefont {A.}~\bibnamefont
  {Carrington}}, \bibinfo {author} {\bibfnamefont {P.}~\bibnamefont
  {Fournier}}, \bibinfo {author} {\bibfnamefont {R.~L.}\ \bibnamefont
  {Greene}}, \bibinfo {author} {\bibfnamefont {P.}~\bibnamefont {Guptasarma}},
  \bibinfo {author} {\bibfnamefont {D.~G.}\ \bibnamefont {Hinks}},\ and\
  \bibinfo {author} {\bibfnamefont {A.~R.}\ \bibnamefont {Banks}},\ }\href@noop
  {} {\bibfield  {journal} {\bibinfo  {journal} {Appl. Phys. Lett.}\ }\textbf
  {\bibinfo {volume} {77}},\ \bibinfo {pages} {4202} (\bibinfo {year}
  {2000}{\natexlab{b}})}\BibitemShut {NoStop}%
\bibitem [{\citenamefont {Prozorov}\ and\ \citenamefont
  {Kogan}(2018)}]{Demag2018}%
  \BibitemOpen
  \bibfield  {author} {\bibinfo {author} {\bibfnamefont {R.}~\bibnamefont
  {Prozorov}}\ and\ \bibinfo {author} {\bibfnamefont {V.~G.}\ \bibnamefont
  {Kogan}},\ }\href {https://doi.org/10.1103/PhysRevApplied.10.014030}
  {\bibfield  {journal} {\bibinfo  {journal} {Phys. Rev. Applied}\ }\textbf
  {\bibinfo {volume} {10}},\ \bibinfo {pages} {014030} (\bibinfo {year}
  {2018})}\BibitemShut {NoStop}%
\bibitem [{\citenamefont {Chandrasekhar}\ and\ \citenamefont
  {Einzel}(1993)}]{Chandrasekhar1993}%
  \BibitemOpen
  \bibfield  {author} {\bibinfo {author} {\bibfnamefont {B.~S.}\ \bibnamefont
  {Chandrasekhar}}\ and\ \bibinfo {author} {\bibfnamefont {D.}~\bibnamefont
  {Einzel}},\ }\href {https://doi.org/https://doi.org/10.1002/andp.19935050604}
  {\bibfield  {journal} {\bibinfo  {journal} {Annalen der Physik}\ }\textbf
  {\bibinfo {volume} {505}},\ \bibinfo {pages} {535} (\bibinfo {year}
  {1993})}\BibitemShut {NoStop}%
\bibitem [{\citenamefont {Einzel}(2003)}]{Einzel2003}%
  \BibitemOpen
  \bibfield  {author} {\bibinfo {author} {\bibfnamefont {D.}~\bibnamefont
  {Einzel}},\ }\href {https://doi.org/10.1023/A:1022872911344} {\bibfield
  {journal} {\bibinfo  {journal} {Journal of Low Temperature Physics}\ }\textbf
  {\bibinfo {volume} {131}},\ \bibinfo {pages} {1} (\bibinfo {year}
  {2003})}\BibitemShut {NoStop}%
\bibitem [{\citenamefont {Prozorov}\ and\ \citenamefont
  {Giannetta}(2006)}]{Prozorov2006}%
  \BibitemOpen
  \bibfield  {author} {\bibinfo {author} {\bibfnamefont {R.}~\bibnamefont
  {Prozorov}}\ and\ \bibinfo {author} {\bibfnamefont {R.~W.}\ \bibnamefont
  {Giannetta}},\ }\href@noop {} {\bibfield  {journal} {\bibinfo  {journal}
  {Superc. Sci. Technol.}\ }\textbf {\bibinfo {volume} {19}},\ \bibinfo {pages}
  {R41} (\bibinfo {year} {2006})}\BibitemShut {NoStop}%
\bibitem [{\citenamefont {Gorter}\ and\ \citenamefont
  {Casimir}(1934{\natexlab{a}})}]{Gorter1934}%
  \BibitemOpen
  \bibfield  {author} {\bibinfo {author} {\bibfnamefont {C.}~\bibnamefont
  {Gorter}}\ and\ \bibinfo {author} {\bibfnamefont {H.}~\bibnamefont
  {Casimir}},\ }\href {https://doi.org/10.1016/s0031-8914(34)90037-9}
  {\bibfield  {journal} {\bibinfo  {journal} {Physica}\ }\textbf {\bibinfo
  {volume} {1}},\ \bibinfo {pages} {306} (\bibinfo {year}
  {1934}{\natexlab{a}})}\BibitemShut {NoStop}%
\bibitem [{\citenamefont {Gorter}\ and\ \citenamefont
  {Casimir}(1934{\natexlab{b}})}]{Gorter1934a}%
  \BibitemOpen
  \bibfield  {author} {\bibinfo {author} {\bibfnamefont {C.~J.}\ \bibnamefont
  {Gorter}}\ and\ \bibinfo {author} {\bibfnamefont {H.}~\bibnamefont
  {Casimir}},\ }\href@noop {} {\bibfield  {journal} {\bibinfo  {journal} {Z.
  tech. Phys}\ }\textbf {\bibinfo {volume} {15}},\ \bibinfo {pages} {539}
  (\bibinfo {year} {1934}{\natexlab{b}})}\BibitemShut {NoStop}%
\bibitem [{\citenamefont {Bardeen}(1958)}]{Bardeen1958}%
  \BibitemOpen
  \bibfield  {author} {\bibinfo {author} {\bibfnamefont {J.}~\bibnamefont
  {Bardeen}},\ }\href {https://doi.org/10.1103/physrevlett.1.399} {\bibfield
  {journal} {\bibinfo  {journal} {Physical Review Letters}\ }\textbf {\bibinfo
  {volume} {1}},\ \bibinfo {pages} {399} (\bibinfo {year} {1958})}\BibitemShut
  {NoStop}%
\bibitem [{\citenamefont {Prozorov}(2021)}]{Prozorov2021}%
  \BibitemOpen
  \bibfield  {author} {\bibinfo {author} {\bibfnamefont {R.}~\bibnamefont
  {Prozorov}},\ }\href {https://doi.org/10.1103/physrevapplied.16.024014}
  {\bibfield  {journal} {\bibinfo  {journal} {Physical Review Applied}\
  }\textbf {\bibinfo {volume} {16}},\ \bibinfo {pages} {024014} (\bibinfo
  {year} {2021})}\BibitemShut {NoStop}%
\bibitem [{\citenamefont {Carrington}(2011)}]{Carrington2011}%
  \BibitemOpen
  \bibfield  {author} {\bibinfo {author} {\bibfnamefont {A.}~\bibnamefont
  {Carrington}},\ }\href
  {https://doi.org/https://doi.org/10.1016/j.crhy.2011.03.001} {\bibfield
  {journal} {\bibinfo  {journal} {Comptes Rendus Physique}\ }\textbf {\bibinfo
  {volume} {12}},\ \bibinfo {pages} {502} (\bibinfo {year} {2011})},\ \bibinfo
  {note} {superconductivity of strongly correlated systems}\BibitemShut
  {NoStop}%
\bibitem [{\citenamefont {Giannetta}\ \emph {et~al.}(2022)\citenamefont
  {Giannetta}, \citenamefont {Carrington},\ and\ \citenamefont
  {Prozorov}}]{Giannetta2022}%
  \BibitemOpen
  \bibfield  {author} {\bibinfo {author} {\bibfnamefont {R.}~\bibnamefont
  {Giannetta}}, \bibinfo {author} {\bibfnamefont {A.}~\bibnamefont
  {Carrington}},\ and\ \bibinfo {author} {\bibfnamefont {R.}~\bibnamefont
  {Prozorov}},\ }\href {https://doi.org/10.1007/s10909-021-02626-3} {\bibfield
  {journal} {\bibinfo  {journal} {Journal of Low Temperature Physics}\ }\textbf
  {\bibinfo {volume} {208}},\ \bibinfo {pages} {119} (\bibinfo {year}
  {2022})}\BibitemShut {NoStop}%
\bibitem [{\citenamefont {Prozorov}\ and\ \citenamefont
  {Kogan}(2011)}]{Prozorov2011}%
  \BibitemOpen
  \bibfield  {author} {\bibinfo {author} {\bibfnamefont {R.}~\bibnamefont
  {Prozorov}}\ and\ \bibinfo {author} {\bibfnamefont {V.~G.}\ \bibnamefont
  {Kogan}},\ }\href {http://stacks.iop.org/0034-4885/74/i=12/a=124505}
  {\bibfield  {journal} {\bibinfo  {journal} {Reports on Progress in Physics}\
  }\textbf {\bibinfo {volume} {74}},\ \bibinfo {pages} {124505} (\bibinfo
  {year} {2011})}\BibitemShut {NoStop}%
\bibitem [{\citenamefont {Donovan}\ \emph {et~al.}(1993)\citenamefont
  {Donovan}, \citenamefont {Klein}, \citenamefont {Dressel}, \citenamefont
  {Holczer},\ and\ \citenamefont {Grüner}}]{Donovan1993}%
  \BibitemOpen
  \bibfield  {author} {\bibinfo {author} {\bibfnamefont {S.}~\bibnamefont
  {Donovan}}, \bibinfo {author} {\bibfnamefont {O.}~\bibnamefont {Klein}},
  \bibinfo {author} {\bibfnamefont {M.}~\bibnamefont {Dressel}}, \bibinfo
  {author} {\bibfnamefont {K.}~\bibnamefont {Holczer}},\ and\ \bibinfo {author}
  {\bibfnamefont {G.}~\bibnamefont {Grüner}},\ }\href
  {https://doi.org/10.1007/bf02086217} {\bibfield  {journal} {\bibinfo
  {journal} {International Journal of Infrared and Millimeter Waves}\ }\textbf
  {\bibinfo {volume} {14}},\ \bibinfo {pages} {2459} (\bibinfo {year}
  {1993})}\BibitemShut {NoStop}%
\bibitem [{\citenamefont {Dressel}\ \emph {et~al.}(1993)\citenamefont
  {Dressel}, \citenamefont {Klein}, \citenamefont {Donovan},\ and\
  \citenamefont {Grüner}}]{Dressel1993}%
  \BibitemOpen
  \bibfield  {author} {\bibinfo {author} {\bibfnamefont {M.}~\bibnamefont
  {Dressel}}, \bibinfo {author} {\bibfnamefont {O.}~\bibnamefont {Klein}},
  \bibinfo {author} {\bibfnamefont {S.}~\bibnamefont {Donovan}},\ and\ \bibinfo
  {author} {\bibfnamefont {G.}~\bibnamefont {Grüner}},\ }\href
  {https://doi.org/10.1007/bf02086218} {\bibfield  {journal} {\bibinfo
  {journal} {International Journal of Infrared and Millimeter Waves}\ }\textbf
  {\bibinfo {volume} {14}},\ \bibinfo {pages} {2489} (\bibinfo {year}
  {1993})}\BibitemShut {NoStop}%
\bibitem [{\citenamefont {Klein}\ \emph {et~al.}(1993)\citenamefont {Klein},
  \citenamefont {Donovan}, \citenamefont {Dressel},\ and\ \citenamefont
  {Grüner}}]{Klein1993}%
  \BibitemOpen
  \bibfield  {author} {\bibinfo {author} {\bibfnamefont {O.}~\bibnamefont
  {Klein}}, \bibinfo {author} {\bibfnamefont {S.}~\bibnamefont {Donovan}},
  \bibinfo {author} {\bibfnamefont {M.}~\bibnamefont {Dressel}},\ and\ \bibinfo
  {author} {\bibfnamefont {G.}~\bibnamefont {Grüner}},\ }\href
  {https://doi.org/10.1007/bf02086216} {\bibfield  {journal} {\bibinfo
  {journal} {International Journal of Infrared and Millimeter Waves}\ }\textbf
  {\bibinfo {volume} {14}},\ \bibinfo {pages} {2423} (\bibinfo {year}
  {1993})}\BibitemShut {NoStop}%
\bibitem [{\citenamefont {Nikolo}(1995)}]{Martin1995}%
  \BibitemOpen
  \bibfield  {author} {\bibinfo {author} {\bibfnamefont {M.}~\bibnamefont
  {Nikolo}},\ }\href {https://doi.org/10.1119/1.17770} {\bibfield  {journal}
  {\bibinfo  {journal} {American Journal of Physics}\ }\textbf {\bibinfo
  {volume} {63}},\ \bibinfo {pages} {57} (\bibinfo {year} {1995})}\BibitemShut
  {NoStop}%
\bibitem [{\citenamefont {Topping}\ and\ \citenamefont
  {Blundell}(2018)}]{Topping2018}%
  \BibitemOpen
  \bibfield  {author} {\bibinfo {author} {\bibfnamefont {C.~V.}\ \bibnamefont
  {Topping}}\ and\ \bibinfo {author} {\bibfnamefont {S.~J.}\ \bibnamefont
  {Blundell}},\ }\href {https://doi.org/10.1088/1361-648x/aaed96} {\bibfield
  {journal} {\bibinfo  {journal} {Journal of Physics: Condensed Matter}\
  }\textbf {\bibinfo {volume} {31}},\ \bibinfo {pages} {013001} (\bibinfo
  {year} {2018})}\BibitemShut {NoStop}%
\bibitem [{\citenamefont {Rutgers}(1934)}]{Rutgers1934}%
  \BibitemOpen
  \bibfield  {author} {\bibinfo {author} {\bibfnamefont {A.}~\bibnamefont
  {Rutgers}},\ }\href {https://doi.org/10.1016/s0031-8914(34)80300-x}
  {\bibfield  {journal} {\bibinfo  {journal} {Physica}\ }\textbf {\bibinfo
  {volume} {1}},\ \bibinfo {pages} {1055} (\bibinfo {year} {1934})}\BibitemShut
  {NoStop}%
\bibitem [{\citenamefont {Kim}\ \emph {et~al.}(2013)\citenamefont {Kim},
  \citenamefont {Kogan}, \citenamefont {Cho}, \citenamefont {Tanatar},\ and\
  \citenamefont {Prozorov}}]{Kim2013}%
  \BibitemOpen
  \bibfield  {author} {\bibinfo {author} {\bibfnamefont {H.}~\bibnamefont
  {Kim}}, \bibinfo {author} {\bibfnamefont {V.~G.}\ \bibnamefont {Kogan}},
  \bibinfo {author} {\bibfnamefont {K.}~\bibnamefont {Cho}}, \bibinfo {author}
  {\bibfnamefont {M.~A.}\ \bibnamefont {Tanatar}},\ and\ \bibinfo {author}
  {\bibfnamefont {R.}~\bibnamefont {Prozorov}},\ }\href
  {https://doi.org/10.1103/PhysRevB.87.214518} {\bibfield  {journal} {\bibinfo
  {journal} {Phys. Rev. B}\ }\textbf {\bibinfo {volume} {87}},\ \bibinfo
  {pages} {214518} (\bibinfo {year} {2013})}\BibitemShut {NoStop}%
\end{thebibliography}
%

\end{document}